\def\astroph{1}                              

\ifnum\astroph=0                             
\documentclass[12pt,preprint]{aastex}        
\else                                        
\documentclass[onecolumn]{emulateapj}
\usepackage{apjfonts}
\fi

\usepackage{epsfig,graphicx,natbib}
\usepackage{amsmath}
\newcommand{\beq}{\begin{equation}}
\newcommand{\eeq}{\end{equation}}
\newcommand{\bea}{\begin{eqnarray}}
\newcommand{\eea}{\end{eqnarray}}
\newcommand{\rem}[1]{ }

\newcommand{\hn}{{\bf\hat n}} 	
\newcommand{\hv}{{\bf\hat v}}
\newcommand{\hs}{{\bf\hat s}}

\newcommand{\tn}{\hat n} 	
\newcommand{\tv}{\hat v}

\begin{document}
\shorttitle{Radiation from sub-Larmor scale magnetic fields}
\shortauthors{Medvedev, Frederiksen, Haugb\o lle, Nordlund}

\title{Radiation signatures of sub-Larmor scale magnetic fields}
\author{Mikhail V. Medvedev$^{1,2,3}$, Jacob Trier Frederiksen$^2$, Troels Haugb\o lle$^2$, \AA ke Nordlund$^2$}
\affil{$^1$Institute for Advanced Study, School of Natural Sciences, Princeton, NJ 08540}
\affil{$^2$Niels Bohr International Academy, Blegdamsvej 17, DK-2199 K\o benhavn \O, Denmark}
\altaffiltext{3}{Also at: Department of Physics and Astronomy, University of Kansas, Lawrence, KS 66045 and Institute for Nuclear Fusion, RRC ``Kurchatov Institute'', Moscow 123182, Russia}

\begin{abstract}

Spontaneous rapid growth of strong magnetic fields is rather ubiquitous in high-energy density environments ranging from astrophysical sources (e.g., gamma-ray bursts and relativistic shocks), to reconnection, to laser-plasma interaction laboratory experiments, where they are produced by kinetic streaming instabilities of the Weibel type. Relativistic electrons propagating through these sub-Larmor-scale magnetic fields radiate in the jitter regime, in which the anisotropy of the magnetic fields and the particle distribution have a strong effect on the produced radiation. Here we develop the general theory of jitter radiation, which includes (i) anisotropic magnetic fields and electron velocity distributions, (ii) the effects of trapped electrons and (iii) extends the description to large deflection angles of radiating particles thus establishing a cross-over between the classical jitter and synchrotron regimes. Our results are in remarkable agreement with the radiation spectra obtained from particle-in-cell simulations of the classical Weibel instability. Particularly interesting is the onset of the field growth, when the transient hard synchrotron-violating spectra are common as a result of the dominant role of the trapped population. This effect can serve as a distinct observational signature of the violent field growth in astrophysical sources and lab experiments. It is also interesting that a system with small-scale fields tends to evolve toward the small-angle jitter regime, which can, under certain conditions, dominate the overall emission of a source.
\end{abstract}

\section{Introduction}

There is a lore that a relativistic particle of charge $e$ and a Lorentz factor $\gamma$ moving through a magnetic field $B$ produces synchrotron radiation, whose spectrum peaks at $\omega_s\sim (eB/mc)\gamma^2$, has an asymptotic $\omega^{1/3}$ dependence below the peak and falls off exponentially at higher frequencies (it makes a second power-law for an isotropic ensemble of particles having a power-law distribution in energy). This is often true, but not always. If the field is inhomogeneous on scales comparable or smaller then the particle Larmor radius, $\lambda_B\lesssim R_L\sim\gamma m c^2/e\langle B\rangle$, the produced radiation spectrum may be far different from synchrotron. 

Gamma-ray bursts, supernovae shocks, relativistic pulsar winds and shocks, relativistic jets from quasars and active galactic nuclei, magnetic reconnection sites, plasmas produced by high-intensity lasers --- they all are the high-energy density environments where conditions are favorable for the spontaneous magnetic field production. The field generation via the Weibel instability \citep{Weibel59,Fried59,Silva+03,WA} or its modifications \citep{D+06,Bret+08,Bret09,Jacob+08} has been predicted to occur in astrophysical shocks with low ambient magnetic field and rare particle collisions, e.g., in gamma-ray burst and large-scale structure shocks \citep{ML99,MSK06,MZ09}. It has been observed in numerical simulations of relativistic non-magnetized shocks \citep{Nish+04,Jacob+04,Spitk08,Keshet+09,Nish+09}, nonrelativistic shocks unmagnetized and weakly magnetized shocks \citep{KT08,KT10}, cosmic rays interacting with a pre-shock medium \citep{Jacek+10}, magnetic reconnection in electron-positron relativistic and non-relativistic plasmas \citep{ZH08,Swisdak+08,Liu+09}, as well as in simulations of and even real laser plasma experiments \citep{Ren+04,T+03}. 

Given such a ubiquity of the process at hand, a natural question to ask is: Are there any observational signatures, which can benchmark the process in astrophysical sources and, if any, what can we learn about the physical conditions there? With the radiation techniques being developed and implemented into numerical codes \citep{Hededal05,Nish+08,SS09,Jacob+2010}, we will soon be able to answer this question in detail \citep[for instance,][showed that PIC simulations can realistically model some astrophysical sources]{MS09}. In order to correctly interpret the results of simulations and observational data, a comprehensive theory of radiation processes in a strong small-scale magnetic turbulence is, therefore, of great demand. 

The effects of small-scale inhomogeneities on radiation emission have been of long-standing theoretical interest \citep[see, for instance,][and many more]{LP53,Migdal54,Migdal56,GS65,LL}. Techniques developed in these papers have further been applied to synchrotron radiation from large-scale homogeneous magnetic fields with a small-scale random field component, as a model of radio emission by cosmic rays in the interstellar medium \citep[see, e.g.,][and references therein]{NT79,BNT80,Toptygin,TF87}. The so-called perturbative approach of radiation emission from random small-scale magnetic fields without a large-scale component has first been discussed as a model of radiation from Weibel-mediated relativistic colisionless shocks of gamma-ray bursts \citep{M00}, where it was referred to as jitter radiation. This approach was generalized in \citep{F06} and further corrected%
\footnote{It is important to note here some problems in \citep{F06}, which are relevant to the present paper. First, it was argued that the spectrum $F_\nu\propto \nu^1$ below the peak, ``valid in the presence of ordered small-scale magnetic field fluctuations, does not occur in the general case of small-scale random magnetic field fluctuations." This statement was shown \citep{M05,M06} to be flawed in that jitter radiation from {\em random} magnetic fluctuations with a fairly general distribution function (not just ``ordered small-scale'' fields) does allow for $\propto \nu^1$ spectra. Moreover, the entire range from $\propto \nu^1$ to $\propto\nu^0$ is allowed for a single electron emission, softer spectra can be expected for some electron energy distributions. Second, another confusing issue is related to the absorption-like $\propto\nu^2$ jitter spectrum. Such a spectrum is due to plasma dispersion and does not occur in the absence thereof, as the reader might incorrectly infer \citep[see][for more discussion]{M05}. Finally, we should note that contrary to the claims, none of their papers treat the jitter regime properly. The approximations made in their analysis apply to the systems with large-scale plus small-scale isotropic turbulent fields, which do not hold for the generic Weibel-like magnetic turbulence.} %
in \citep{M05,M06,MPR09,RPM10}. Because of this and also because \citep{M06} was the first to consider anisotropic magnetic turbulence (e.g., Weibel- or filamentation-instability-generated magnetic field turbulence), we will refer to this paper in the following discussion. We also would like to mention here a recent paper by \citep{RK10}, who developed a new algorithm to compute radiation from small-scale turbulent fields.

In this paper we develop a theory of jitter radiation that accounts for anisotropies of the magnetic field and particle velocity distributions, including a trapped population, and further extend the theory to the large angle jitter regime. Our theoretical findings are tested with dedicated particle-in-cell simulations. Interesting conclusions are presented in the final section.

\section{Theory}

Radiation emitted from magnetic fields with small coherence length is {\em not} synchrotron, regardless of the actual shape of the produced spectrum. In some cases, the spectrum may resemble that of synchrotron, while in others it can be markedly different. What kind of spectrum is produced is, in general, set by how curved the particle paths are (i.e., how large their deflections from a straight line) compared to the relativistic beaming angle $\sim1/\gamma$. In particular, when the deflection angle $\alpha\sim e\langle B\rangle\lambda_B/\gamma m c^2$ is smaller than the beaming angle the particle radiates in the classical jitter regime \citep{M00,M06}, in which the particle's velocity ${\bf v}$ is almost constant, its path is almost straight, ${\bf r=v}t$, and its acceleration ${\bf w\equiv\dot v}$ is random and varies rapidly in time. Qualitatively, in the small-angle jitter regime, i.e., when the jitter parameter  
\beq
\delta_{\rm jitt}=e\langle B\rangle\lambda_B/m c^2=\gamma\lambda_B/R_L
\simeq5.9\times10^{-4}\left(\lambda_B/1~\textrm{cm}\right)\left(\langle B\rangle/1~\textrm{gauss}\right)
\eeq
is small, $\delta_{\rm jitt}\ll1$, the spectrum has a peak at $\omega_j\sim (c/\lambda_B)\gamma^2$. The spectral shape is generally not universal at lower frequencies (it can be flat, $F_\nu\propto\nu^0$, in the isotropic magnetic turbulence but can also be as steep as $F_\nu\propto\nu^1$ in the presence of a strong anisotropy) and is usually a power-law above the peak, whose index is related to that of the spectrum of the magnetic turbulence and/or particle distribution. 

\subsection{Jitter radiation in the small deflection angle regime}
\label{s:j}

Here we generalize the theory of classical jitter radiation. Energy emitted by an accelerated relativistic particle and observed at infinity is given by the Poynting flux, which is easily calculated using Li\'enard-Wiechert (retarded) potentials. One arrives at the familiar expression \citep{LL} for the total energy emitted per unit solid angle $d O$ per unit frequency $d\omega$:
\beq
dW=\frac{e^2}{2\pi c^3}\frac{\left|\hn\times\left[\left(\hn-\hv\beta\right)\times
{\bf w}_{\omega'}\right]\right|^2}{\left(1-\beta\hv\cdot\hn\right)^{4}}\ d O\,\frac{d\omega}{2\pi},
\label{dW}
\eeq
where $\beta=v/c=(1-\gamma^{-2})^{1/2}$, a ``hat'' denotes unit vectors, ${\bf w}_{\omega'}=\int{\bf w}e^{i\omega't}\,dt$ is the Fourier component of the particle acceleration, the frequency in the comoving and observer's frames are related as $\omega'=\omega\left(1-\beta\hv\cdot\hn\right)$, and $\hn$ points toward the observer. Since acceleration in a magnetic field, ${\bf w}=(e/\gamma m c){\bf v\times B}$, is orthogonal to ${\bf v}$, the following holds:
\beq
\left|\hn\times\left[\left(\hn-\hv\beta\right)\times
{\bf w}_{\omega'}\right]\right|^2
=\left|{\bf w}_{\omega'}\right|^2 \left(1-\beta\hv\cdot\hn\right)^2-
\left|\hn\cdot{\bf w}_{\omega'}\right|^2 \gamma^{-2},
\label{term}
\eeq
where we keep the small term $\sim\gamma^{-2}$ because it can be important if a particle velocity distribution is structured at angular scales $\lesssim1/\gamma$ and/or if $\gamma$ is not very large.

So far, no approximations were made. Now we use the standard small-deflection approximation, ${\bf v}=const$, and we let the magnetic field vary, hence the Fourier image of acceleration and its projection onto $\hn$ are $w^\alpha_{\omega'}=(e\beta/\gamma m)\frac{1}{2}e_{\alpha\beta\gamma}(\tv_\beta B^\gamma_{\omega'}-\tv_\gamma B^\beta_{\omega'})$ and $\tn_\alpha w^\alpha_{\omega'}=(e\beta/\gamma m)\frac{1}{2}e_{\alpha\beta\gamma}B^\alpha_{\omega'}(\tn_\beta\tv_\gamma-\tn_\gamma\tv_\beta)$, where $e_{\alpha\beta\gamma}$ is the Lev\'i-Civita tensor and in flat configuration space we do not distinguish between co- and contra-variant components. Now we immediately obtain\footnote{Here we used the identities
$$
e_{\alpha\beta\gamma}e_{\lambda\mu\nu}=
\left|
\begin{array}{ccc}
\delta_{\alpha\lambda} & \delta_{\alpha\mu} & \delta_{\alpha\nu} \\
\delta_{\beta\lambda} & \delta_{\beta\mu} & \delta_{\beta\nu} \\
\delta_{\gamma\lambda} & \delta_{\gamma\mu} & \delta_{\gamma\nu} 
\end{array}
\right|
\textrm{~and~}
e_{\alpha\beta\gamma}e_{\alpha\mu\nu}=
\left|
\begin{array}{cc}
\delta_{\beta\mu} & \delta_{\beta\nu} \\
\delta_{\gamma\mu} & \delta_{\gamma\nu} 
\end{array}
\right|
$$}
\bea
\left|{\bf w}_{\omega'}\right|^2&=&
\left(\delta_{\alpha\beta}-\tv_\alpha\tv_\beta\right) W^{\alpha\beta}_{\omega'} 
\label{w}\\
\left|\hn\cdot{\bf w}_{\omega'}\right|^2&=&
\left[\delta_{\alpha\beta}\left(1-(\hv\cdot\hn)^2\right)-\tv_\alpha\tv_\beta-\tn_\alpha\tn_\beta
+(\hv\cdot\hn)(\tv_\alpha\tn_\beta+\tv_\beta\tn_\alpha)
\right] W^{\alpha\beta}_{\omega'}
\label{nw}
\eea
where $W^{\alpha\beta}_{\omega'}=(e\beta/\gamma m)^2 B_{\omega'}^\alpha B_{\omega'}^{*\beta}$ is the acceleration tensor, ${\bf B_{\omega'}}$ represents the temporal variation of the field along the particle path and $\delta_{\alpha\beta}$ is the Kronecker symbol. In general, the field varies in space and time 
\beq
{\bf B}(t,{\bf r})=(2\pi)^{-4}\int e^{-i(\Omega t -{\bf k \cdot r})}{\bf B}_{\Omega,{\bf k}}\,d\Omega\,d{\bf k}.
\eeq
For a straight path ${\bf r}={\bf r}_0+{\bf v}t$, the field is a function of one independent variable $t$, hence
\bea
{\bf B_{\omega'}}&=&(2\pi)^{-4}\int e^{i\omega' t} dt \left(e^{-i(\Omega t -{\bf k}\cdot{\bf v}t -{\bf k}\cdot{\bf r}_0)}{\bf B}_{\Omega,{\bf k}}\right) d\Omega\,d{\bf k} 
\nonumber \\
&=&(2\pi)^{-3}\int \delta(\omega'-\Omega+{\bf k\cdot v})\, e^{i{\bf k}\cdot{\bf r}_0}{\bf B}_{\Omega,{\bf k}}\, d\Omega\,d{\bf k},
\eea
where we used the identity for the Dirac $\delta$-function: $\int e^{i\xi t}\,dt=2\pi\delta(\xi)$.

So far we dealt with a particular representation of a particle motion through magnetic turbulence. The assumption of ergodicity allows us to relate the ``representative'' history of $B$-field along a path to the spatial average over all possible initial positions ${\bf r}_0$. This is a strong assumption, but it is valid for statistically homogeneous turbulence with no correlation between particles and fields (this is not strictly true for trapped particles, see below). Thus
\bea
\langle B_{\omega'}^\alpha B_{\omega'}^{*\beta}\rangle&=&
(2\pi)^{-6}V^{-1}\int B^\alpha_{\Omega,{\bf k}} B^{*\beta}_{\Omega_1,{\bf k}_1}\,
\delta(\omega'-\Omega+{\bf k\cdot v})\, \delta(\omega'-\Omega_1+{\bf k}_1\cdot {\bf v})\,
e^{i({\bf k}-{\bf k}_1)\cdot{\bf r}_0} d{\bf r}_0
d\Omega d\Omega_1d{\bf k}d{\bf k}_1
\nonumber \\
&=&(2\pi)^{-3}V^{-1}\int B^\alpha_{\Omega,{\bf k}} B^{*\beta}_{\Omega,{\bf k}}\,
\delta(\omega'-\Omega+{\bf k\cdot v})\, d\Omega d{\bf k},
\label{BBomega}
\eea
where $V$ is the volume occupied by the magnetic field and we again used the identity $\int e^{i({\bf k}-{\bf k}_1)\cdot{\bf r}_0} d{\bf r}_0=(2\pi)^3\delta({\bf k}-{\bf k}_1)$. It is trivial to prove via straightforward substitution\footnote{
Indeed, $B^\alpha_{\Omega,{\bf k}} B^{*\beta}_{\Omega,{\bf k}}=\int\! B_\alpha(t',{\bf r}')e^{i(\Omega t' -{\bf k\cdot r}')}dt'd{\bf r}'\, \int\! B^*_\beta(t'',{\bf r}'')e^{-i(\Omega t'' -{\bf k\cdot r}'')}dt''d{\bf r}''= 
\int[\int\! B_\alpha(t',{\bf r}')B_\beta(t'-t,{\bf r}'-{\bf r})dt'd{\bf r}']e^{i(\Omega t-{\bf k\cdot r})}dtd{\bf r}
=TV \int \langle B_\alpha B_\beta(t,{\bf r})\rangle\, e^{i(\Omega t- {\bf k\cdot r})} dt d{\bf r}$.}
that the spectral tensor $B^\alpha_{\Omega,{\bf k}} B^{*\beta}_{\Omega,{\bf k}}$ is just a Fourier image of the two-point autocorrelation tensor of the field,
$\langle B_\alpha B_\beta(t,{\bf r})\rangle \equiv T^{-1}V^{-1}\int B_\alpha(t',{\bf r}') B_\beta(t'-t,{\bf r'-r})\,dt'd{\bf r}'$, that is :
\beq
B^\alpha_{\Omega,{\bf k}} B^{*\beta}_{\Omega,{\bf k}}=TV \int
\langle B_\alpha B_\beta(t,{\bf r})\rangle e^{i(\Omega t- {\bf k\cdot r})} dt\,d{\bf r},
\eeq
where $T$ is the duration of an observation. Eqs. (\ref{dW})--(\ref{nw}) and (\ref{BBomega}) completely determine the jitter radiation spectrum of a single particle in the small-deflection regime for an arbitrary field distribution.

\begin{figure}[t!]
\center\includegraphics[angle = 0, width = 0.45\columnwidth]{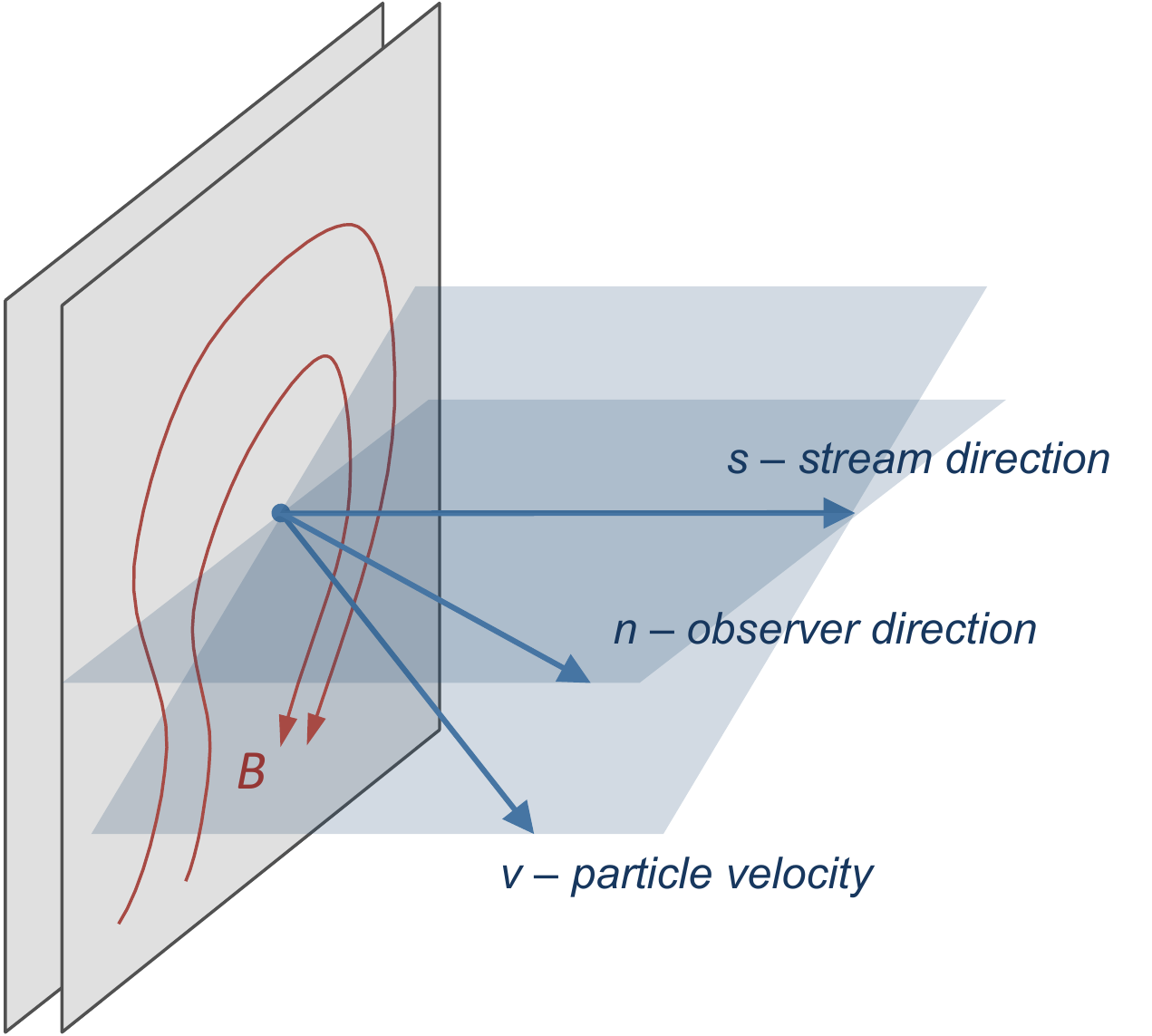}
\caption{A cartoon showing the assumed planar distribution of the magnetic field typical of the Weibel turbulence and the triad of the unit vectors: $\hs,\ \hn,\ \hv$. }
\label{triad}
\end{figure}

When magnetic field is generated by a Weibel-type instability, particles bunch into localized streams -- current filaments -- thus determining a preferred direction, $\hs$, so that the magnetic field is predominantly in the plane orthogonal to  $\hs$, as is illustrated in Figure \ref{triad}, so 
\beq
B^\alpha_{\Omega,{\bf k}} B^{*\beta}_{\Omega,{\bf k}}=\left| B_{\Omega,{\bf k}}\right|^2(\delta_{\alpha\beta}-s_\alpha s_\beta),
\eeq
where $\left| B_{\Omega,{\bf k}}\right|^2$ is the standard scalar spectrum of the field. The acceleration tensor becomes 
\beq
W_{\omega'}^{\alpha\beta}=(\delta_{\alpha\beta}-s_\alpha s_\beta)\left|W_{\omega'}\right|^2, 
\label{W}
\eeq
where $\left|W_{\omega'}\right|^2$ is the scalar frequency spectrum of the particle acceleration. Finally, the spectral energy emitted by an ensemble of particles with a homogeneous anisotropic velocity distribution, $F({\bf v})$, is
\bea
\frac{dW^{\rm ens}}{d O\,d\omega}&=&\frac{e^2}{(2\pi)^2 c^3}\int\left[\frac{1+(\hs\cdot\hv)^2}{\left(1-\beta\hv\cdot\hn\right)^{2}}-\frac{1}{\gamma^2}\frac{(\hs\cdot\hv)^2+(\hs\cdot\hn)^2-2(\hv\cdot\hn)(\hs\cdot\hv)(\hs\cdot\hn)}{\left(1-\beta\hv\cdot\hn\right)^{4}}\right]
\nonumber\\
& &\times\left[\left(\frac{e\beta}{\gamma m}\right)^2\frac{1}{(2\pi)^{3}V}\int \left| B_{\Omega,{\bf k}}\right|^2\,\delta\!\left(\omega(1-\beta\hv\cdot\hn)-\Omega+{\bf k\cdot v}\right)\, d\Omega d{\bf k}\right] F({\bf v})\, d{\bf v} .
\label{dWmain}
\eea
The terms in this expression with septuple integration have clear physical meanings. The two terms in the first square brackets are the geometric factors coming from the product of $(\delta_{\alpha\beta}-s_\alpha s_\beta)$ in Eq. (\ref{W}) with the tensors in front of $W_{\omega'}^{\alpha\beta}$ in Eqs. (\ref{w}), (\ref{nw}) respectively, the term in the second square brackets is simply the particle acceleration spectrum $\left|W_{\omega'}\right|^2$ and the outer integral weighted with the particle distribution function sums up the contributions of all particles in the system.

It is worthwhile to note that (i) the radiation spectrum of a single particle is proportional to the spectrum of the particle accelerations 
\beq
dW/d O\, d\omega\propto |{\bf w}_{\omega'}|^2
\eeq
and, hence, is a `probe' of the magnetic field structure along its path and (ii) the radiation is strongly beamed in the direction of the particle motion: 
\beq
dW/d O\propto (1-\beta(\hn\cdot\hv))^{-3}\propto(1+(\gamma\vartheta)^2)^{-3}.
\eeq
The latter expression is valid for $\gamma\gg1$ and a small angle $\vartheta$ between ${\bf v}$ and the line of sight. If the magnetic field is static ($\Omega=0$) and its spatial spectrum has a peak at a characteristic coherence scale, $k_B\sim\lambda_B^{-1}$, then the single electron emissivity in the small angle regime is peaked at the frequency
\beq
\omega_{j,{\rm sm}}\sim k_Bc\gamma^2\sim(c/\lambda_B)\gamma^2.
\eeq

\subsection{Jitter radiation from the trapped population}

The Weibel instability \citep{Weibel59} is driven by anisotropy of the particle distribution $F({\bf v})$, which has been interpreted as the instability of streaming particles \citep{Fried59}. An infinitesimal transverse modulation of the stream density (i.e., the current density) results in transverse magnetic fields, which pinch the particles into filamentary structures and therefore enhance the initial perturbation. The filamentary distribution of particle streams is maintained by the self-generated magnetic fields. These stream particles are essentially trapped in the filaments for a long time and, hence, the ergodicity assumption used in the derivation of jitter radiation in Section \ref{s:j} fails for them, so they are not accounted for by Eq. (\ref{dWmain}). These particles oscillate in filaments and radiate. Depending on the oscillation amplitude $\theta_m$ they emit radiation in the small-angle jitter regime if $\theta_m<1/\gamma$ and in the large-angle jitter regime otherwise. However, in both cases one needs to know exact particle trajectories in order to calculate radiation. This is not possible for a generic magnetic field distribution. We illustrate this in the following simplified example of a two-dimensional straight filament.  

\begin{figure}[b]
\center\includegraphics[angle = 0, width = 0.5\columnwidth]{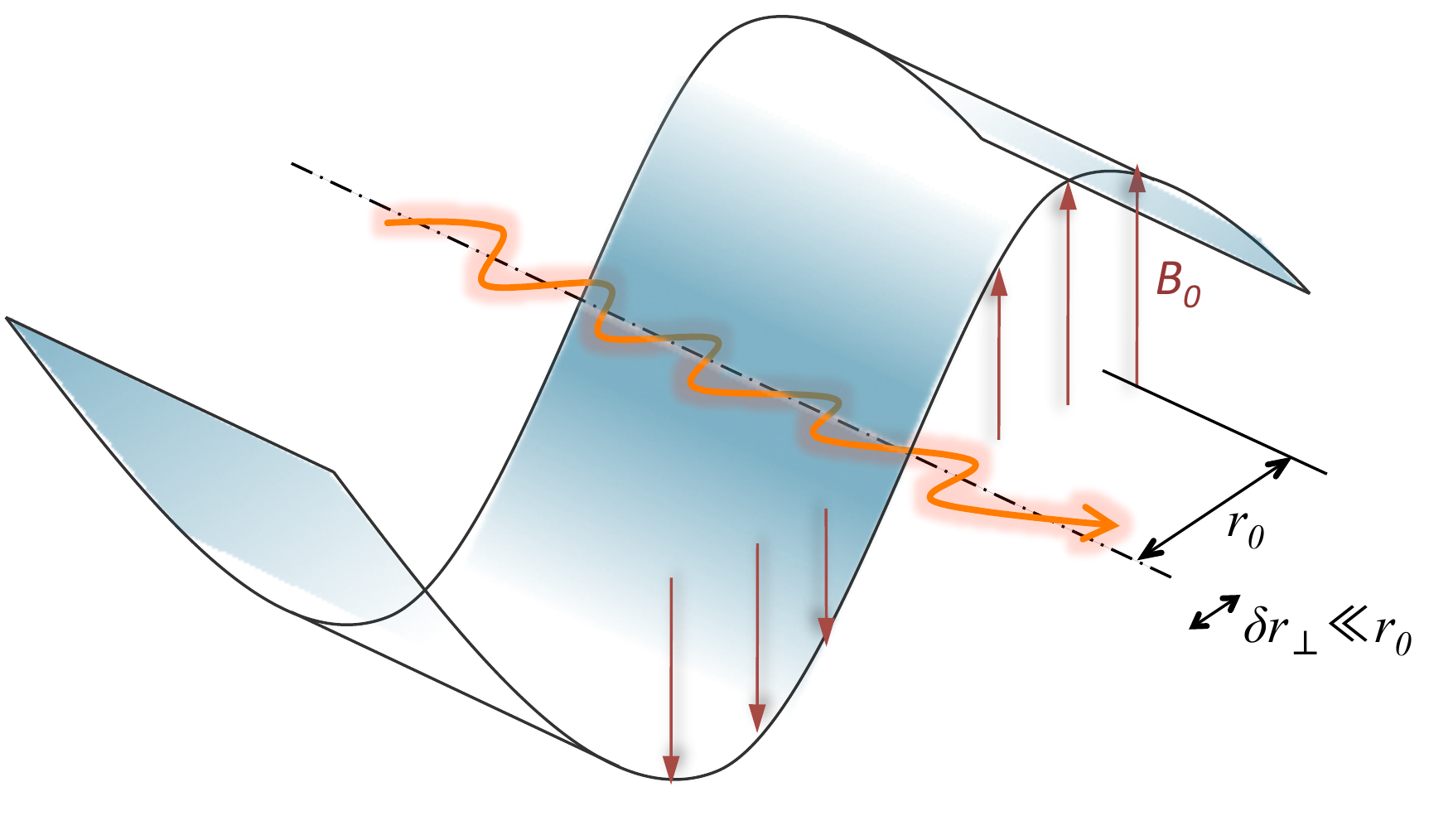}
\caption{A cartoon showing a two-dimensional model of a radiating particle trapped in a filament with the magnetic field shown with vectors and the curved envelope. The color gradient illustrates the distribution of a current density. The $\delta r_\perp$ is the particle oscillation amplitude, the filament size is $r_0$ and $B_0=B(r_0)$. }
\label{trapped}
\end{figure}

If there is translational invariance along the filament, the parallel component of the generalized momentum of a particle is conserved: $P_\|=\gamma m v_\|-(e/c)A_\|(r_\perp)=const$, where $A_\|(r_\perp)$ is the only non-zero component of the vector potential and it is a function of the transverse coordinate only. Since ${\bf E}=0$ in the system, the particle energy is also conserved, therefore $v=const$. Because of the axial symmetry, the magnetic field is zero on the axis $r_\perp=0$, so we can take $A_\|(0)=0$. Then, it is straightforward to obtain the equation
\beq
\frac{dr_\perp}{dt}=\left[v^2-\left(v_\|(0)+\frac{e}{\gamma m}\,A_\|(r_\perp)\right)^2\right]^{1/2},
\eeq
which can be solved only in quadrature to obtain the particle trajectory as $t=t(r_\perp)$; here $v_\|(0)$ is the parallel velocity on the axis of the filament. Even the turning points, $r_\perp=r_t$, can only be found implicitly: $v_\|(0)+({e}/{\gamma m})\,A_\|(r_t)=v$.

This model can be simplified further, as shown in Figure \ref{trapped}. For a uniform distribution of current near the filament axis, the magnetic field is a linear function of coordinates $B(r_\perp)=(r_\perp/r_0)B_0$, which holds for small displacements $r_\perp\ll r_0$, where $r_0$ is the transverse size of a filament and $B_0$ is the field on its ``surface''. The equation of motion $\dot {\bf p}=(e/c){\bf v}\times{\bf B}$ for small amplitudes and $v\sim c$ reads:
\beq
\ddot r_\perp\simeq (e B_0/\gamma m)(r_\perp/r_0).
\eeq
This is the equation of a harmonic oscillator. The streaming particles are trapped in filaments and oscillate with the characteristic bounce frequency
\beq
\Omega_b\simeq \left(eB_0/\gamma m r_0\right)^{1/2}.
\label{Omegab}
\eeq
Due to the oscillatory motion, the magnetic field in the frame comoving with the particle varies in time as $B(t)\simeq B_0\left(r_t\sin(\Omega_b t)/r_0\right)\propto \sin(\Omega_b t)$. For small oscillation amplitudes\footnote{
This is a more restrictive condition than $\delta r_\perp\ll r_0$, for which  $\theta_m$ can also be larger than $1/\gamma$, depending on the value of $r_0$.}
 $\theta_m\sim v_\perp/v\ll1/\gamma$, the particle path can still be approximated as a straight line, so Eq. (\ref{dWmain}) can be used. To account for the bounce motion, the magnetic field spectrum can be approximately taken in the form\footnote{Alternatively, one can set $\Omega=0$ (static field) and modify the $k_\|$-spectrum of the magnetic field to have a sharp peak at $k_\|\sim k_b=\Omega_b/c$.}
\beq
\left| B_{\Omega,{\bf k}}\right|^2\sim\left| B_{\bf k}\right|^2\delta(\Omega-\Omega_b).
\eeq
If there is a distribution of filaments of various sizes $r_{0,i}$ and strength $B_{0,i}$ in the system, the $\delta$-function shall be substituted with the actual distribution of bounce frequencies, $f_b(\Omega)$. The observed radiation from the trapped population is peaked at the frequency
\beq
\omega_{j,{\rm tr}}\sim\Omega_b\gamma^2.
\label{trapjitt}
\eeq

\subsection{Spectra in the small angle jitter regime}

\begin{figure}[b]
\center\includegraphics[angle = 0, width = 0.5\columnwidth]{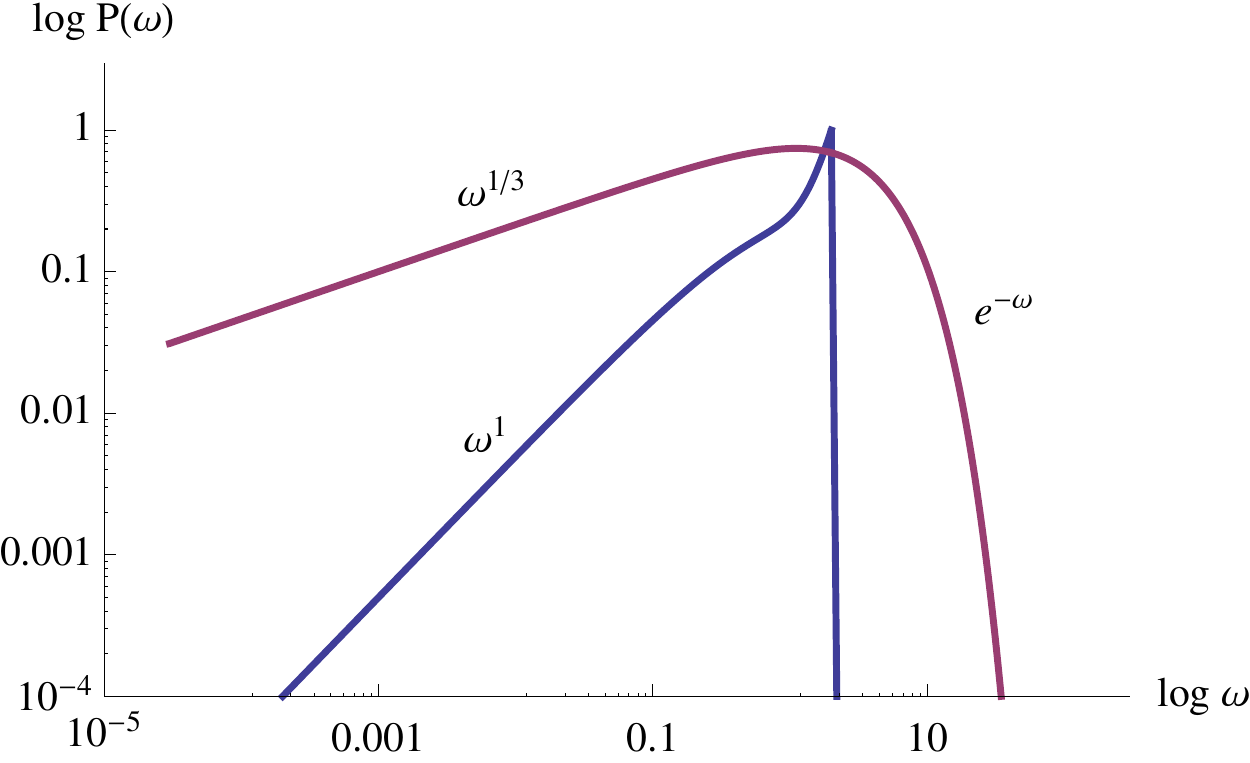}
\caption{Radiation `Green's functions'. The blue sharp curve is the single-electron angle-averaged emission spectrum (equivalent to the ensemble-averaged spectrum from monoenergetic isotropic electrons) from the turbulent magnetic field with a separable $\delta$-function spectrum. The red smooth curve is the single-electron radiation spectrum in the synchrotron regime, shown for comparison. The frequencies are normalized to the respective characteristic frequencies, i.e., $\omega_j=\omega_s=1$. }
\label{greenf}
\end{figure}

As we have mentioned earlier, the small-angle jitter spectra are not universal: they depend on the turbulent magnetic field spectrum and its anisotropy, in addition to the electron distribution which can also be anisotropic. Figure \ref{greenf} shows `Green's function' of jitter radiation --- the single-electron angle-averaged emission spectrum (equivalent to the ensemble-averaged spectrum from monoenergetic randomly moving electrons) from the turbulent magnetic field with a separable $\delta$-function spectrum, i.e.,  $|B_{\Omega,{\bf k}}|^2=\delta(k_x-k_B)\delta(k_y-k_B)\delta(k_z-k_B)$ with $k_B\sim1/\lambda_B$. Other radiation spectra are generally obtained via convolutions with the field spectra and electron distributions. For comparison, we also show the single-electron synchrotron spectrum.

\begin{figure}
\center\includegraphics[angle = 0, width = 0.5\columnwidth]{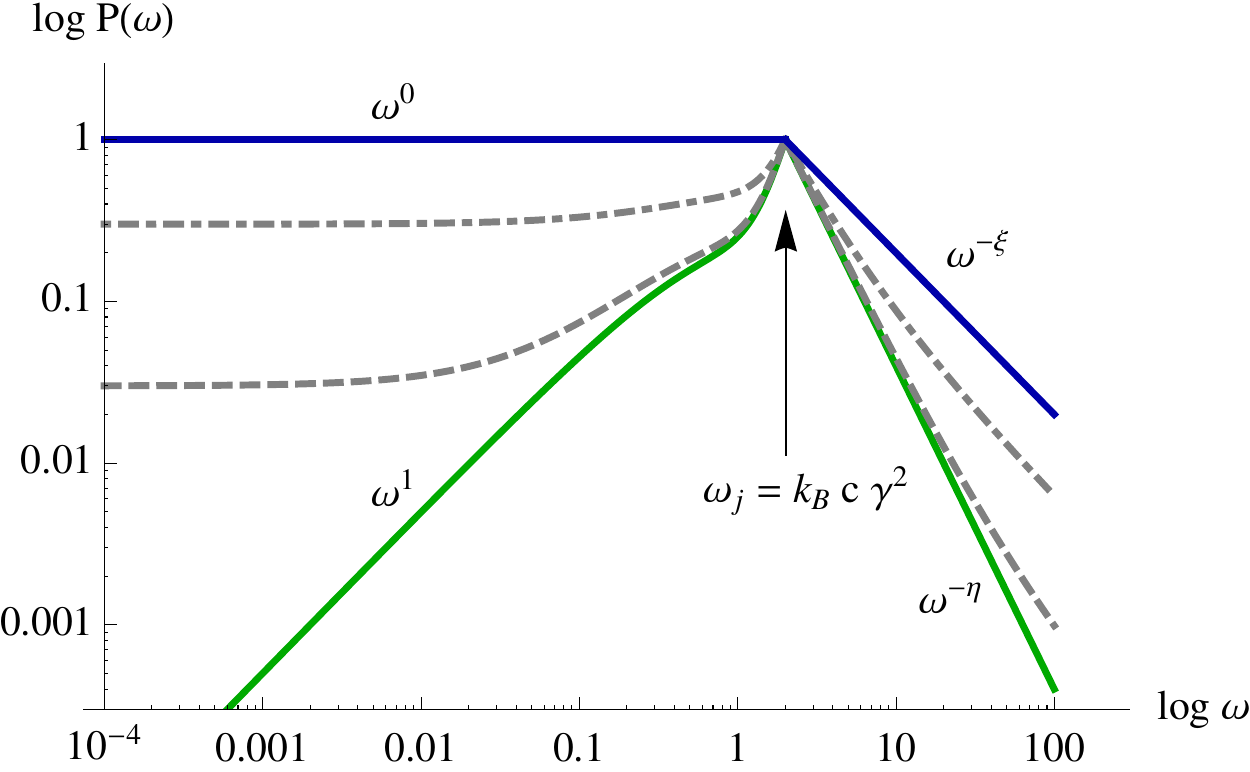}
\caption{Characteristic small-angle jitter radiation spectra from small-scale magnetic turbulence. The green sharply-peaked curve is the spectrum emitted in the direction of the strong anisotropy of the field. This is also the spectrum of the particles trapped in magnetic filaments. The blue double-power-law spectrum is emitted in the direction, orthogonal to the anisotropy direction. This is also the spectrum from the isotropic magnetic turbulence. The grey dashed curves represent spectra emitted at some intermediate angles with respect to the filaments.}
\label{jittset}
\end{figure}

Weibel turbulence is generally anisotropic with the streaming direction and the transverse plane dynamics being decoupled. This is confirmed with PIC simulations, see below. The field spectrum can be represented as a separable function $|B_{\Omega,{\bf k}}|^2=f_\bot(k_\bot)f_z(k_z)$ with $k_\bot^2=k_x^2+k_y^2$. Here we assume  $f_\bot(k_\bot)\propto k_\bot^{-\beta},\ k_\bot>k_B$ and $f_z(k_z)\propto k_z^{-\beta},\ k_z>k_B$. The spectrum of radiation in this case is angle-dependent \citep[see][for a detailed analysis]{M06,RPM10}. The spectrum observed along the filament direction ($z$-direction) is shown in Figure \ref{jittset} with the green sharply-peaked curve resembling the jitter spectrum in Figure \ref{greenf}. The low-energy power-law index is 1 and the high-energy index $\eta$ is related to the field distribution index as $\eta=\alpha$. Note that such harder-than-synchrotron $\omega^1$ spectrum is also produced by trapped particles regardless of the $f_z$-spectrum. The spectrum of radiation observed in the transverse direction (orthogonal to the filaments) is shown with the blue curve. The spectrum is flat below the jitter frequency and is a power-law above. The high-energy index is also related to the field spectrum: $\xi=\beta-1$. The spectra observed at intermediate angles are shown with dashed grey curves.

In the course of non-linear evolution, Weibel turbulence becomes more isotropic. Such `aged' turbulence is expected, for example, in the region far behind a Weibel-mediated shock. The field spectrum becomes isotropic, $|B_{\Omega,{\bf k}}|^2=f(k)$ with $k^2=k_x^2+k_y^2+k_z^2$. We can assume here, for simplicity, that $f(k)\propto k^{-\beta},\ k>k_B$. In this case, the spectrum is also represented by a broken power-law blue curve, which is flat below the break and falls off with the index $\xi=\beta-2$.

\subsection{Jitter radiation in the large deflection angle regime}
\label{large}

When the deflection angle of a particle, $\theta_d\sim(\delta p)/p\sim p_\perp/p$, is greater than $1/\gamma$ (that is, $\delta_{\rm jitt}>1$), the radiation spectrum is determined by the geometry of the particle trajectory. The peak frequency depends on how fast the beaming cone sweeps through the line of sight, as in synchrotron radiation. The power of lower frequency harmonics, however, depends on the global structure of the path, such as the deflection angle $\theta_d$ or the oscillation amplitude $\theta_m$ of a trapped particle. 

\begin{figure}[b!]
\center\includegraphics[angle = 0, width = 0.6\columnwidth]{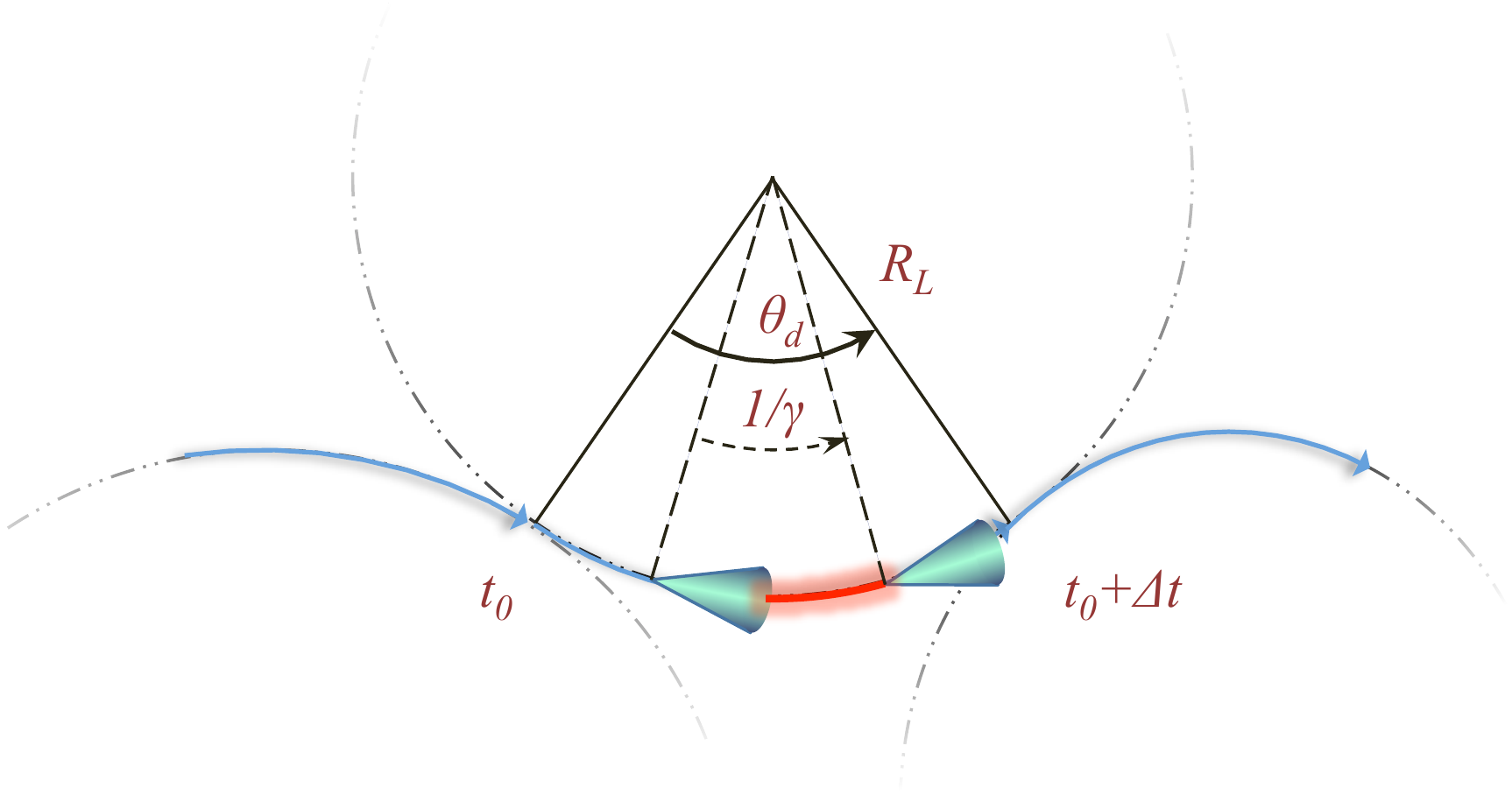}
\caption{Cartoon showing the piecewise circular decomposition of a particle trajectory. For each segment of a certain curvature (Larmor) radius, $R_L$, and the deflection angle, $\theta_d$, radiation is produced at times between $t_0$ and $t_0+\Delta t$. Radiation is beamed directly toward an observer on the right within a highlighted $\sim1/\gamma$ part of the segment.}
\label{largejitter}
\end{figure}

The radiation spectrum can be qualitatively obtained as follows \citep[for non-perturbative approach, see, e.g.,][]{Toptygin}. A generic particle trajectory can approximately be represented as a smoothly joined set of circular segments (arches) of a certain curvature (Larmor) radius, $R_L\sim \gamma mc^2/e\langle B\rangle$, and an angular extent, $\theta_d$, as is illustrated in Figure \ref{largejitter}. Within each segment, $\theta<\theta_d$, the radiation harmonics are constructed coherently in the way similar to the standard synchrotron radiation. For a harmonics constructed over the angle $\theta$, the radiation formation length is $l\sim R_L\theta$. Radiation emitted over this length is observed in the lab frame over the time interval $\Delta t\sim l/c\bar\gamma^2$, where $\bar\gamma$ is the mean Lorentz factor of the particle toward the observer, which is smaller than $\gamma$ because the transverse motion of a particle is also relativistic: $\gamma_\perp\sim\gamma\theta$, hence $\bar\gamma\sim\gamma/\gamma_\perp\sim1/\theta$. The observed frequency is $\omega_\theta\sim1/\Delta t\sim\theta^3c/R_L\sim(\gamma\theta)^{-3}\omega_0\gamma^2$, where $\omega_0=c/R_L$ is the fundamental (Larmor) frequency. The peak (synchrotron) frequency corresponds to the angle $\theta\sim1/\gamma$ (i.e., when the observer is within the radiation cone) $\omega_s=\omega_{1/\gamma}\sim\omega_0\gamma^2$, hence $\omega_\theta\sim(\gamma\theta)^{-3}\omega_s$. The radiation spectrum resembles synchrotron spectrum with $\omega^{1/3}$ law below the peak down to the break jitter frequency $\omega_j\sim\delta_{\rm jitt}^{-3}\omega_s$, where $\delta_{\rm jitt}=(\gamma\theta_d)^{-1}$ by definition. At larger angular scales, $\theta>\theta_d$, the field is effectively incoherent, therefore radiation occurs in the small-angle jitter regime with the effective coherence scale $\lambda_B\sim R_L\theta_d$. 

\begin{figure}
\center\includegraphics[angle = 0, width = 0.5\columnwidth]{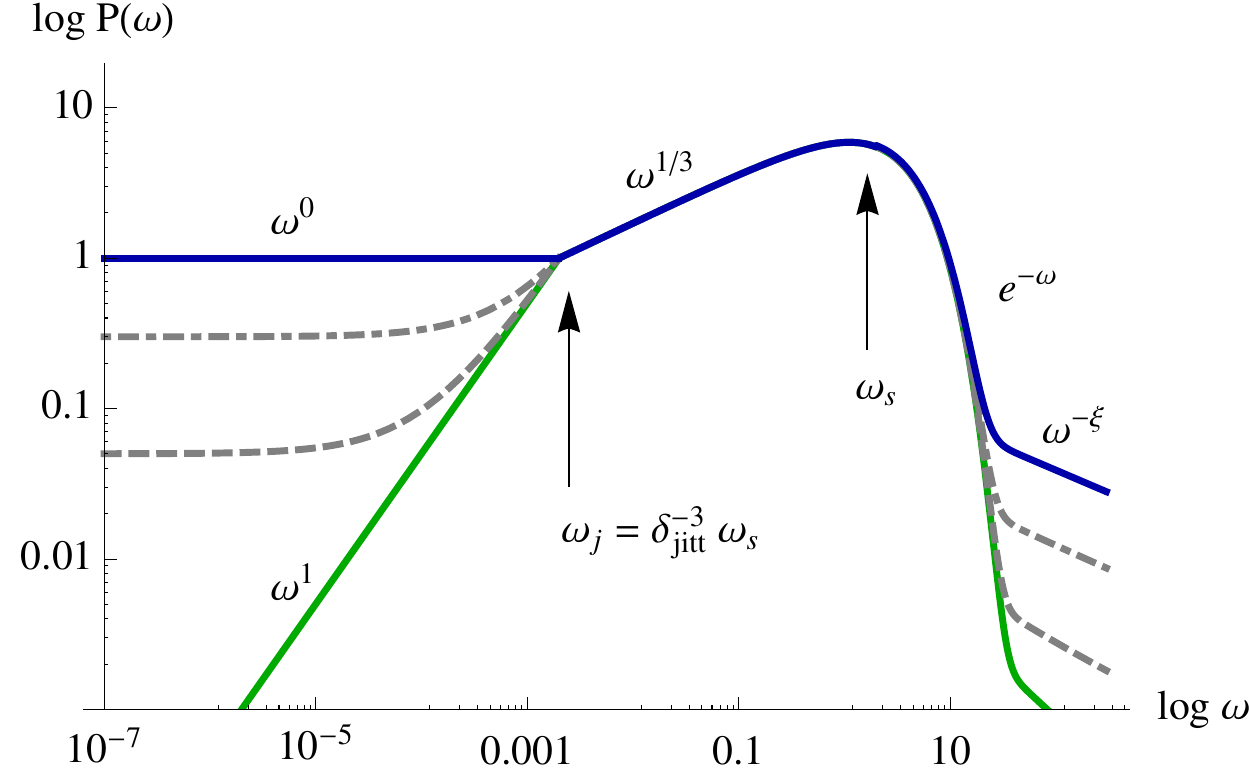}
\caption{Characteristic radiation spectra in the large-angle jitter regime. The spectrum is synchrotron-like above the jitter break and resembles the small-angle jitter spectra below the break, cf. Figure \ref{jittset}. Note that the jitter high-energy power-law can show up above the synchrotron exponential decay at sufficiently large energies. The jitter break frequency is $\omega_j=\delta_{\rm jitt}^{-3}\omega_s$, where $\delta_{\rm jitt}>1$ is the jitter parameter and $\omega_s$ is the synchrotron frequency.}
\vskip1em
\label{jittsynchset}
\end{figure}

To summarize this subsection, the jitter radiation in the large angle regime corresponds to intermediate values of the jitter parameter: $1<\delta_{\rm jitt}<\gamma$. The radiation spectrum is shown in Figure \ref{jittsynchset}. It resembles the synchrotron spectrum,  $\propto\omega^{1/3}\exp(-\omega/\omega_s)$, above the break frequency
\beq
\omega_j\sim\delta_{\rm jitt}^{-3}\omega_s\sim \left(\gamma\lambda_B/R_L\right)^{-3}\left(e\langle B\rangle/mc\right)\gamma^2\sim\left(c/\lambda_B\right)^3\left(e\langle B\rangle/\gamma mc\right)^{-2}
\label{largejitt}
\eeq
and is similar to the small angle jitter spectrum below this break.  At sufficiently large energies, the jitter high-energy power-law shows up above the synchrotron exponential decay. Although the small angle jitter  spectrum is not universal, it is likely flat $\propto\omega^0$ because the field is generally more isotropic at larger scales and trapping of particles is less efficient, but spectra as hard as $\propto\omega^1$ cannot be excluded and may occur under certain conditions. Note that the break frequency, $\omega_j$, depends on the product $\langle B\rangle\lambda_B$ via $\delta_{\rm jitt}$ and the spectral peak $\omega_s$ which measures $B$ (they both proportional to $\gamma^2$ as well), so one can, in principle, determine the magnetic field correlation length $\lambda_B$ from the spectrum alone.
Finally, the case of $\delta_{\rm jitt}>\gamma$ corresponds to synchrotron radiation.

\section{Comparison with spectra from PIC simulations}

\subsection{PIC setup and results}

The PIC simulations of the relativistic filamentation (Weibel) instability \citep{Weibel59,Fried59} have been performed and radiation from the system has simultaneously been obtained \citep[see,][for technical details of the simulations]{Jacob+2010}. The simulations represent the classical Weibel instability with two equal density charge-neutral electron-positron and electron-ion plasma streams. Both two-dimensional and three dimensional setups with various initial bulk Lorentz factors, $\Gamma$, in the range of 2 to 15 were used. In three-dimensional electron-positron pair plasma simulations reported here (an exhaustive description is presented by \citealp{Jacob+2010}) the simulation box of $500^3$ cells with periodic boundary conditions and with the resolution of 10 cells per  the relativistic skin length, $\delta_e=\omega_{pe}/c=\left(4\pi e^2 n/\Gamma m_e c^2\right)^{1/2}$, was used. All physical spatial and temporal scales are expressed in units of the skin length and the plasma time, $\omega_{pe}^{-1}$, respectively. Spectra are collected \textit{in situ} during runtime using the \textsc{PhotonPlasma} code developed at the Niels Bohr Institute \citep{Haugboelle05,Hededal05}. In the simulations reported here, the total number of particles in the simulation box was $\sim10^{10}$; however, the spectra were collected from about $N\sim10^6$ particles and sub-cycling $\Delta t_{\rm rad} \leq 10^{-1}\Delta t_{\rm PIC}$ with linear trajectory interpolation was used to resolve high radiation frequencies. The spectra were calculated coherently from the retarded electric fields of the ensemble of emitting particles
\beq
\frac{dW^{\rm ens}}{d\omega\,d\eta} \propto \left| \int^{t_1}_{t_0} \sum_j^N E_{{\rm ret},j} e^{i\omega\phi'} dt' \right|^2,
\eeq
where the phase is $\phi' \equiv t'-\hn\cdot{\bf r}_0(t')/c$ and we neglected self-absorption, Rasin and other plasma effects. Thus, the obtained spectra represent `time-resolved' radiation emitted within the time interval $\{t_0,t_1\}$. Snapshots of the system (particles, fields, radiation) are saved every two plasma times. Here we report the `head-on' case, when the radiation is emitted along the initial streaming direction. The results for oblique and edge-on cases are reported elsewhere \citep{Jacob+2010}.

\begin{figure}[t!]
\vskip1em
\center\includegraphics[angle = 0, width = 0.8\columnwidth]{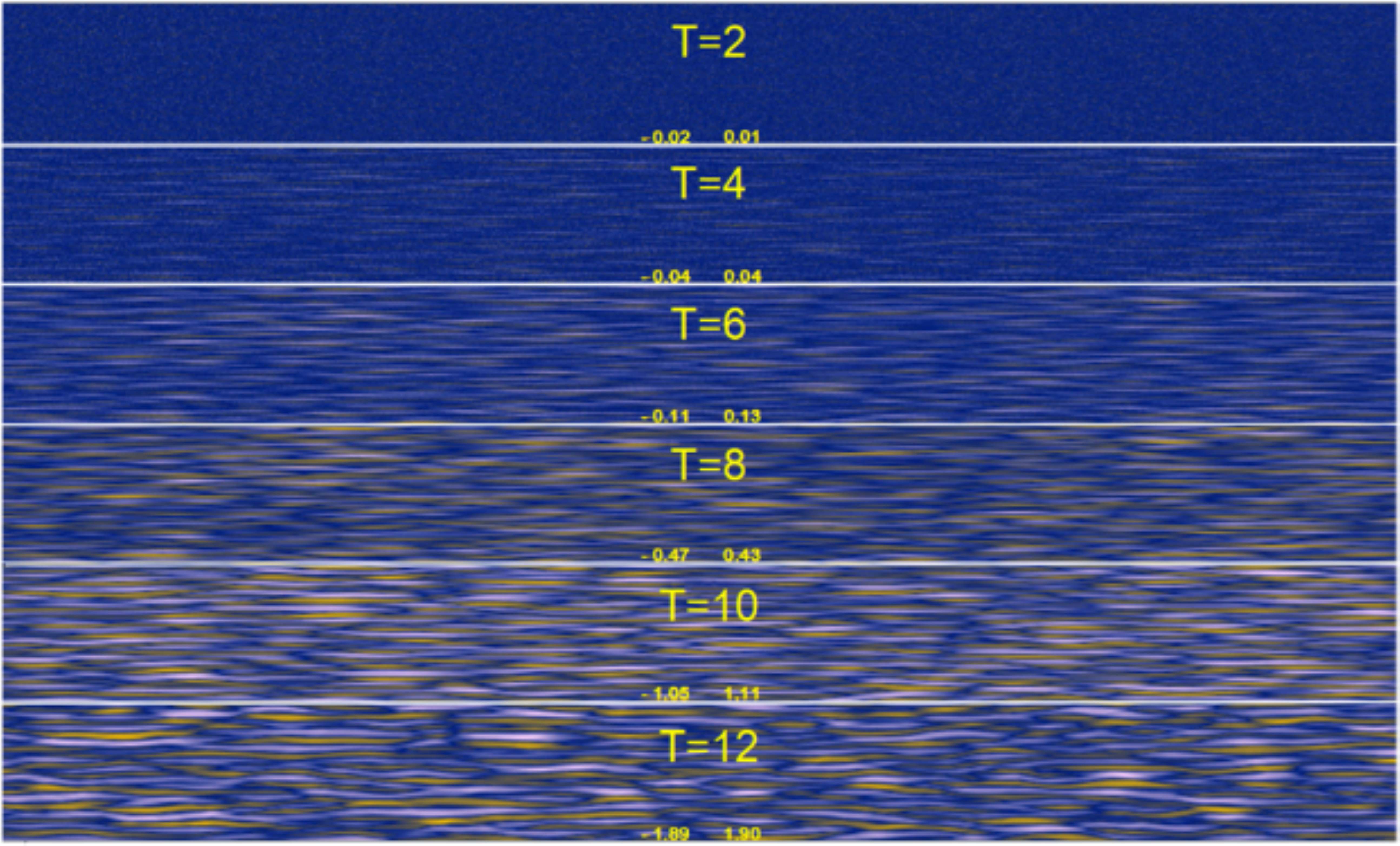}
\caption{The magnetic field distribution in 2D simulations, for illustration purpose only. The simulation box is $2000\times40$ skin lengths, snapshots are at times $t\simeq2,\ 4,\ 6,\ 8,\ 10,\ 12~\omega_{pe}^{-1}$. The field magnitude (with the polarity: into and out of the plane) is coded with white and yellow colors. The pair of numbers at the bottom of each panel shows min and max values of the field in simulation units. Saturation occurs around $t\sim12$ in this particular run; before this the filaments grow in amplitude but not in transverse size.  Note that at saturation, the local field amplitude keeps increasing for some time due to the nonlinear evolution and pinching of the filaments. The overall field energy density is decreasing, however, due to the decreasing filling factor. }
\label{Bmap}
\end{figure}
\begin{figure}
\center\includegraphics[angle = 0, width = 0.5\columnwidth]{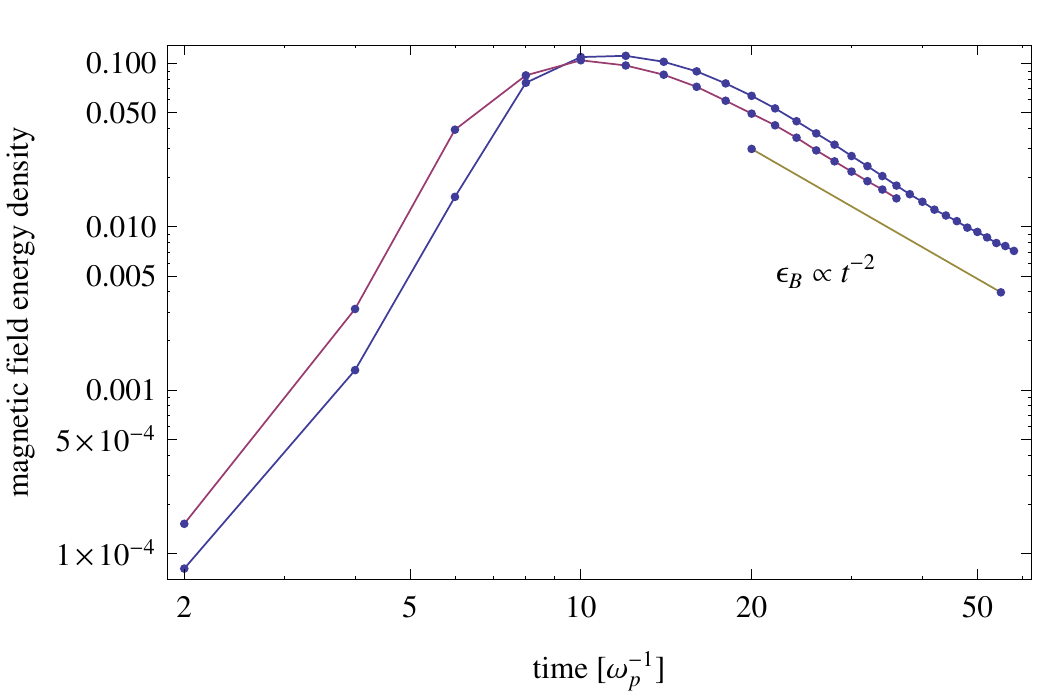}
\caption{The magnetic field energy density normalized by the initial kinetic energy density of the streams, $\epsilon_B$, versus time in realistic 3D simulations for two runs with $\Gamma=2$ (blue) and $\Gamma=10$ (red). At late times, $\epsilon_B\propto t^{-2}$. }
\label{epsB}
\end{figure}

The PIC simulations cover the initial exponential growth of the magnetic field, saturation of the Weibel instability and the nonlinear evolution and mergers of current filaments \citep[the merger model is developed in][]{M+05,Shvets+09} when the magnetic field gradually decays. The distribution of the magnetic filaments during the instability growth and saturation is illustrated in Figure \ref{Bmap} (this is the illustration only; the conditions of the scientific runs are different, see above). The evolution of the field strength at different stages of the instability is shown in Figure \ref{epsB}, where 
\beq
\epsilon_B=B^2/\left[8\pi \Gamma(\Gamma-1)nm_e c^2\right],
\eeq 
the magnetic field energy normalized by the total initial kinetic energy in the system, is plotted. The field strength reaches maximum $\sim10\%$ at about saturation, $t\sim10$. After saturation, the field strength averaged over the simulation box decreases as $B(t)\propto t^{-1}$.  Note, however, that the local field amplitude can still increase for some time after saturation due to the nonlinear evolution and pinching of the filaments, but the overall field energy density falls because of the decreasing filling factor. 

\begin{figure}[t!]
\center\includegraphics[angle = 0, width = 0.95\columnwidth]{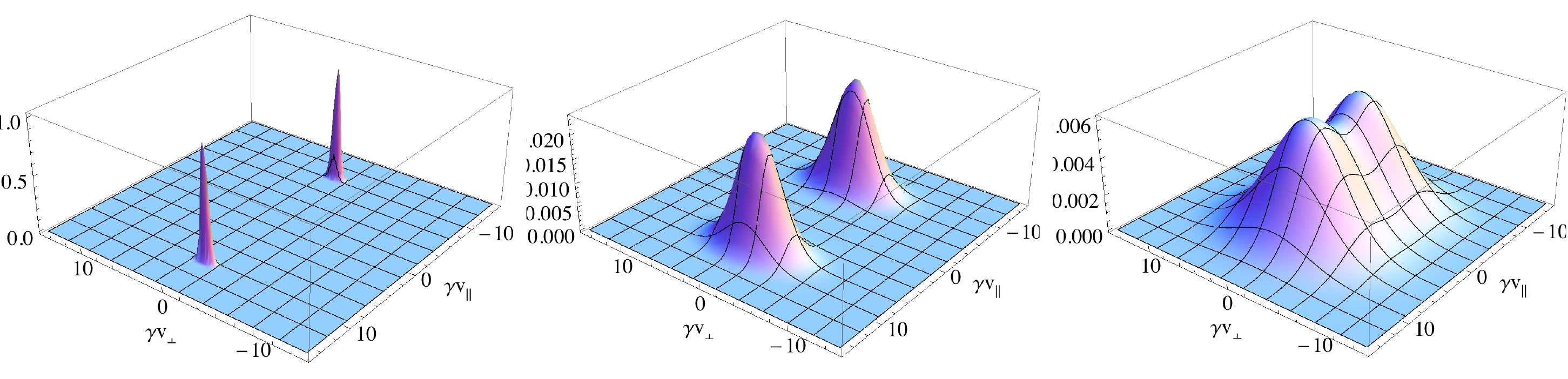} \\ 
\center\includegraphics[angle = 0, width = 0.95\columnwidth]{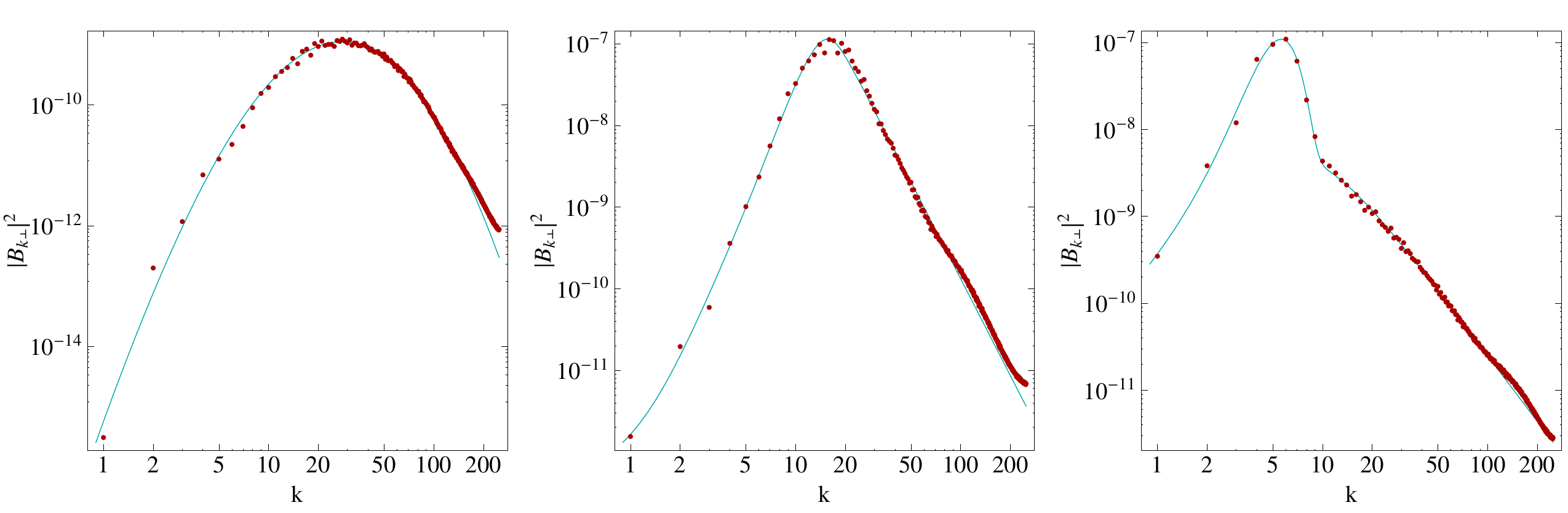} \\
\center\includegraphics[angle = 0, width = 0.95\columnwidth]{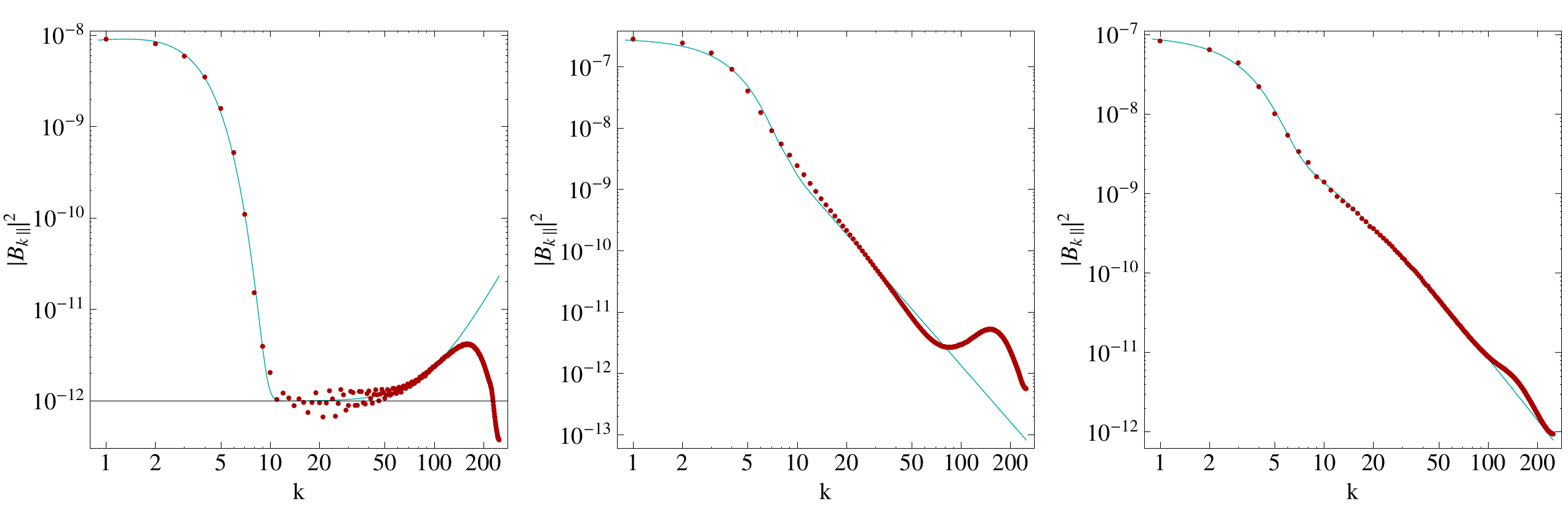}
\caption{A representative sample of snapshots of (top to bottom) the particle distribution, $F(\gamma v_\perp,\gamma v_\|)$, the perpendicular, $|B_{k_\perp}|^2$, and parallel, $|B_{k_\|}|^2$, magnetic field spectra before, at and after saturation (left to right: $t=6,\, 10,\, 30$) for the 3D PIC simulation with $\Gamma=10$. The blue curves are the empirical models, while the red dots are computed directly from the data.}
\label{sample}
\end{figure}

The particle distribution function, $F(\gamma v_\perp,\gamma v_\|)$, and the transverse and parallel (with respect to the streaming direction) magnetic field spectra $|B_{k_\perp}|^2\equiv\langle B_{\perp}({k_\perp})B^*_{\perp}({k_\perp})\rangle$ and $|B_{k_\|}|^2\equiv\langle B_{\perp}({k_\|})B^*_{\perp}({k_\|})\rangle$ are shown in Figure \ref{sample} at representative times: during the linear stage of the instability with exponential growth ($t=6$), at saturation ($t=10$) and the late nonlinear regime ($t=30$). Hereafter, we consider $\Gamma=10$ case only; other cases with relativistic $\Gamma$ are very similar. 

The particle distribution is well described by a multi-Gaussian in $\gamma v_x,\ \gamma v_y,\ \gamma v_z$. It is represented by two shifted Gaussians in the streaming (parallel) direction and a Gaussian in the perpendicular plane. The parallel and transverse temperatures are different and vary in time, so does the average (bulk) Lorentz factor of the interpenetrating streams. 

Concerning the electromagnetic field, the electric fields and parallel magnetic fields are present but very weak (typically, less than a percent), so we only use ${\bf B}_\perp$ in calculations of the field spectra. It was also found that the field spectrum is separable, that is, it can be expressed as a product of two independent functions: $|B_{\bf k}|^2=|B_{k_\perp}|^2|B_{k_\|}|^2$, each being evaluated independently. In order to account for the entire evolution, we used a model of a Gaussian plus a smoothly broken power-law. Since the number of the fit parameters is rather large, some of them (but different) were kept fixed at various times. At early times, the perpendicular spectra, $|B_{k_\perp}|^2$, are fit well by a Gaussian or a double-power-law with a broad transition region whose width quickly reduces with time. At saturation, it is still a double-power-law. After saturation, the peak of the spectrum is much better described by a Gaussian component, which moves toward lower-$k$, leaving behind a single power-law at higher $k$. At late times, the power-law exhibits some curvature (downturn) at small $k$, where it intersects with the Gaussian component. In the beginning, the parallel spectra, $|B_{k_\|}|^2$, are flat (nearly at the noise level) with a broad peak at small $k$ and then they develop a power-law. This peak is broad and is always at the largest scale (smallest $k$) corresponding to the box size. The peak is caused by the ``causality horizon": the filaments at distances greater than the light crossing time are uncorrelated thus producing white noise at small $k$. The peak width is decreasing as $\sim 1/t$, therefore. The bump at large $k$ is likely due to the numerical Cherenkov instability. The power of these Fourier modes is very small and they do not significantly affect radiation spectra (although some signatures of it can be discerned). Overall, except for the linear phase of the instability, the temporal evolution of the parallel spectrum is very modest.

\subsection{PIC and semi-analytic radiation spectra}

Semi-analysical small-angle jitter spectra are obtained from Eq. (\ref{dWmain}) for each PIC snapshot using the available data for the particles $F({\bf v})=F(\gamma v_\perp,\gamma v_\|)$ and fields $\left| B_{{\bf k},\Omega}\right|^2= f_\perp(k_\perp)f_\|(k_\|)f_b(\Omega)$, where $f_\perp(k_\perp)=|B_{k_\perp}|^2,\ f_\|(k_\|)=|B_{k_\|}|^2$ and the bounce frequency distribution is taken heuristically, using Eq. (\ref{Omegab}), as follows. If all the filaments are identical, then the total transverse size of the filament $\sim 2r_0$ corresponds to the peak of the $f_\perp(k_\perp)$ distribution, $2r_0\sim k_\perp^{-1}$. We conjecture that the profile of $f_\perp$, especially near the peak, describes the ensemble of filaments in the system. This is certainly violated at $k_\perp$ greater than the skin scale beyond which no filaments exist. These scales are suppressed with an exponential factor in our model, therefore. Taking into account that $\Omega\sim c/r_0\sim 2c/(2r_0)\sim2k_\perp c$, we have $f_b(\Omega)=\left[ 2\Omega f_\perp(2\Omega)e^{-2\Omega/\omega_p}\Gamma(\Gamma-1)/\gamma^2\right]^{1/4}$. Although not rigorous, this prescription works well, likely because the main effect is due to the presence of a characteristic frequency $\Omega_b$ whereas the exact shape of $f_b(\Omega)$ is of lesser importance. We have also computed the synchrotron spectra for the same plasma parameters, i.e., the spectra one would expect from the large-scale magnetic field of the same strength with the same electron energy distribution. These spectra are similar to the large-angle jitter spectra, except for the low-frequency part, where a shallower spectrum is expected in the jitter regime (see discussion in Section \ref{large}). 

\begin{figure}[t!]
\center\includegraphics[angle = 0, width = 0.49\columnwidth]{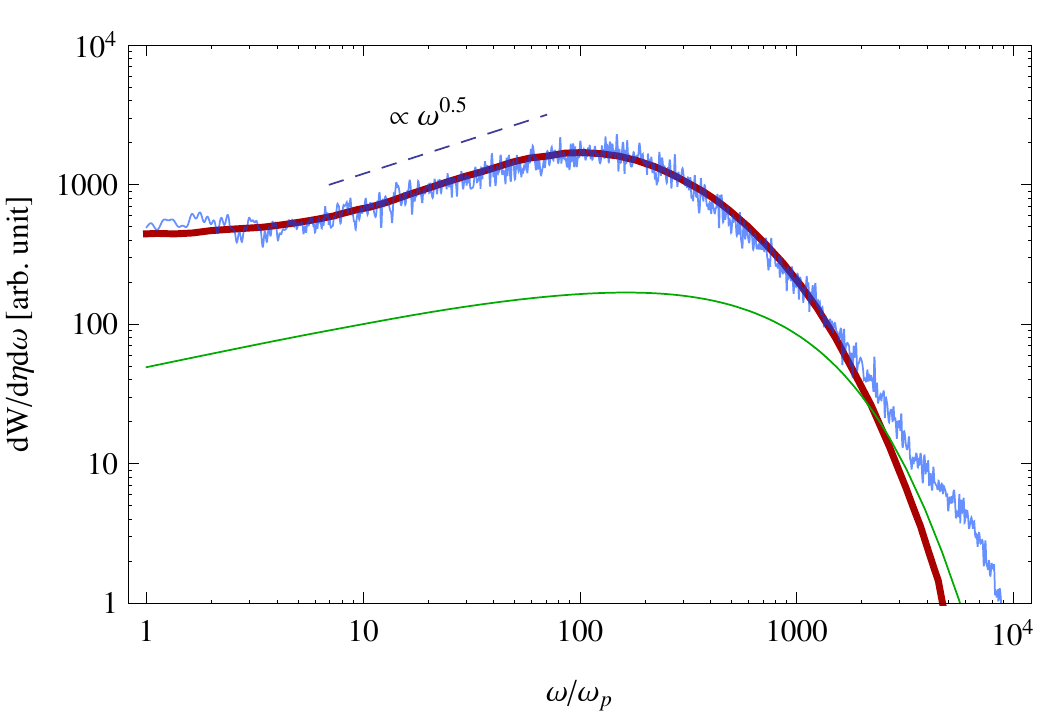}
\includegraphics[angle = 0, width = 0.49\columnwidth]{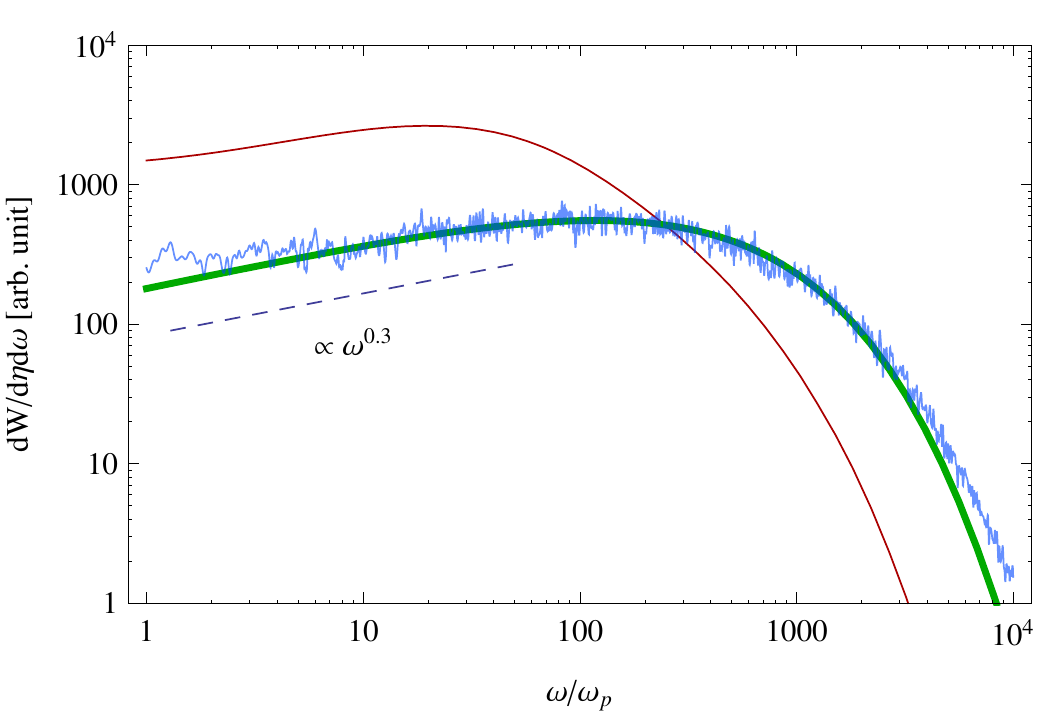} 
\caption{Radiation spectra at early, $t\in\{4,12\}$, (left panel) and late, $t\in\{14,26\}$, times (right panel). The spectra obtained `on-the-flight' from 3D PIC $\Gamma=10$ simulations of the Weibel instability in the $e^+e^-$-pair plasma are the two `noisy' blue curves. The semi-analytic spectra from Eq. \ref{dWmain} are shown in red and the synchrotron spectra are in green. The early time PIC spectra are consistent with the hard (synchrotron violating) small-angle jitter emission and the late time emission is consistent with the large-angle jitter or synchrotron.}
\label{spec}
\end{figure}

The comparison of the PIC and semi-analytic spectra are shown in Fig. \ref{spec} for two time intervals: $t\in\{4,12\}$ which corresponds to the early exponential growth before saturation and $t\in\{14,26\}$ which corresponds to the late nonlinear phase of the filament merger and field decay. The predicted small-angle jitter and classical synchrotron spectra are plotted for comparison. The early-time PIC spectrum is in agreement with the semi-analytic prediction but not with the synchrotron spectrum both in the position of the peak and the overall spectral shape. In particular, (i) the PIC spectrum exhibits synchrotron-violating $\sim\omega^{0.5}$ scaling law below the peak with the subsequent flattening at lower frequencies, (ii) the high-frequency part in the PIC spectrum does not show the near-exponential roll-off (note that the electron distribution is still close to monoenergetic at this time), and (iii) the width of the peak region in the PIC is substantially narrower than it is in the synchrotron spectrum. The late-time PIC spectrum is generally consistent with synchrotron, except at frequencies below $\omega\sim10$, where some flattening becomes evident. 

\begin{figure}[t!]
\center\includegraphics[angle = 0, width = 0.5\columnwidth]{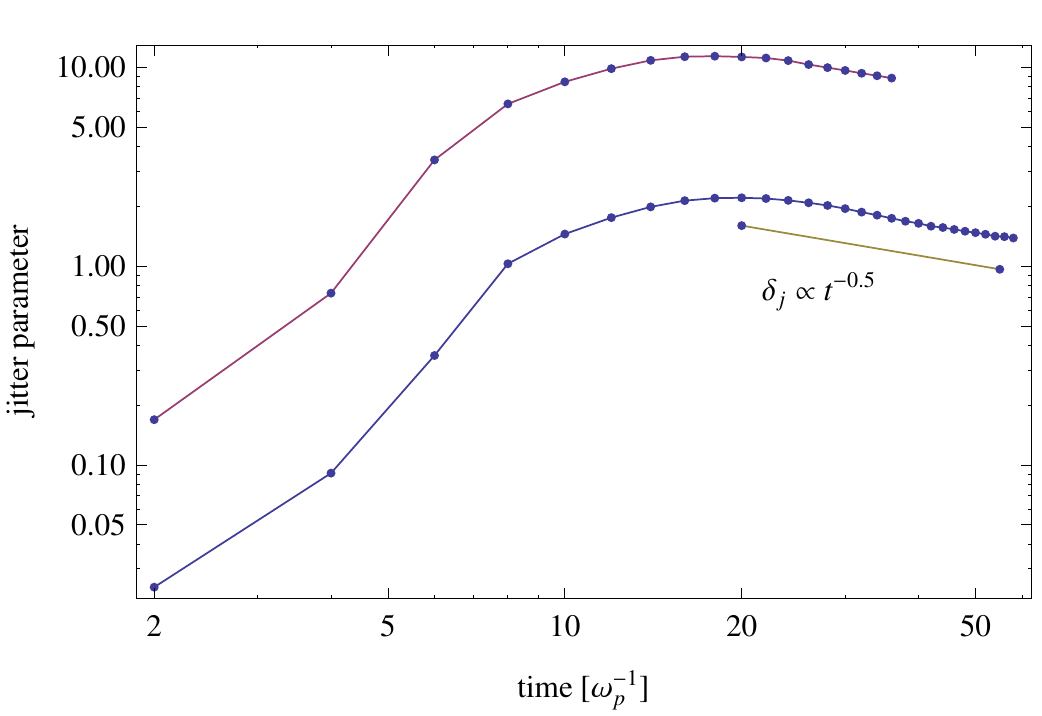}
\caption{Jitter parameter as a function of time for two runs: $\Gamma=10$ (top curve) and $\Gamma=2$ (bottom curve). Asymptotically, $\delta_{\rm jitt}\propto t^{-0.5}$.}
\label{delta}
\end{figure}
\begin{figure}[t!]
\center\includegraphics[angle = 0, width = 0.5\columnwidth]{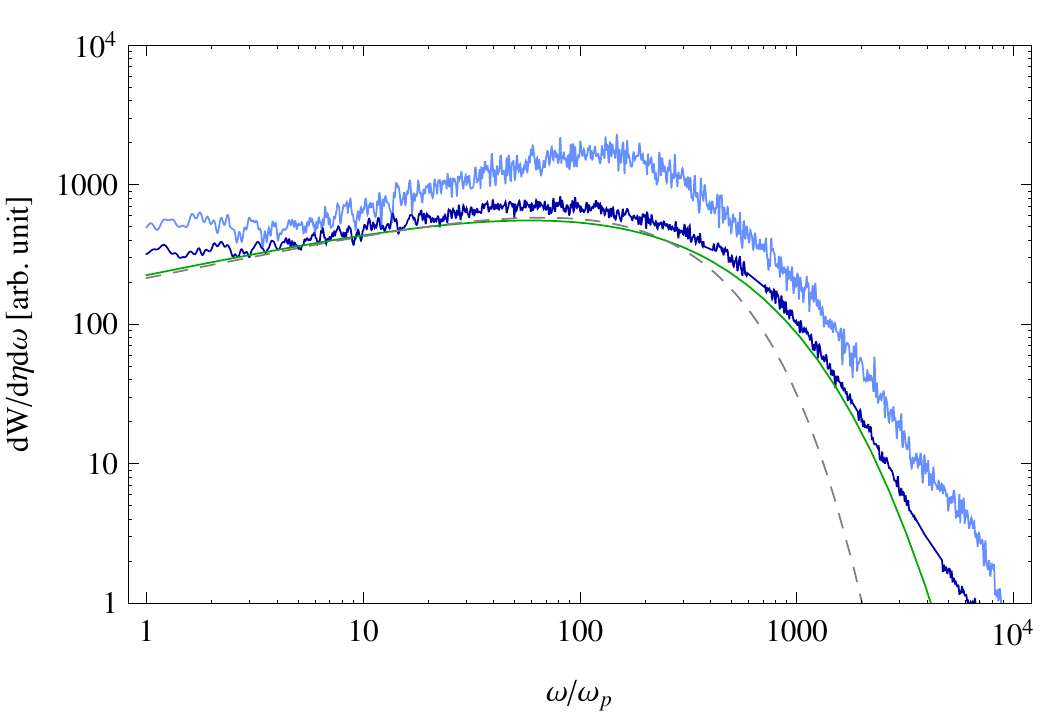}
\caption{The PIC spectrum from the ``frozen field'' simulation  (dark blue) and the original spectrum from Fig. \ref{spec}(left) (light blue) are shown (in both, $\Gamma=10$). The magnetic field structure is nearly identical in both, but the trapped population is practically absent in the ``frozen-field" case. The corresponding synchrotron spectrum for the actual particle distribution (green) and for the monoenergetic electrons (dashed grey) are also shown for comparison. }
\label{frozen}
\end{figure}

The fact that the late-time PIC spectrum is synchrotron-like isn't surprising because $\delta_{\rm jitt}>1$ in the run (see Section \ref{large}), which is evident from Figure \ref{delta}, where we plotted the jitter parameter
\beq
\delta_{\rm jitt}\simeq\left[2\epsilon_B\Gamma(\Gamma-1)\right]^{1/2}(k_{\rm skin}/k_{\perp,\rm max}),
\eeq
where the peak of the $B_{k_\perp}$-spectrum, $k_{\perp,\rm max}$, is found numerically at each time and $k_{\rm skin}$ is the wave-number corresponding to the skin scale. What is particularly interesting is that the early-time PIC spectrum is not consistent with synchrotron at all. The explanation to this is simple. The particle distribution is highly anisotropic at these times: there are still well resolved particle streams which form and are also trapped in the current filaments. For these particles, the size of the magnetic structures is irrelevant. As long as their angular deflections are small enough, they emit in the small-angle regime and produce hard spectra, no matter how large the $\delta_{\rm jitt}$ parameter is. We have confirmed it by using tracer particles in a ``frozen field'' simulation. The full snapshot of particles and fields at $t=8$ was used and in the subsequent simulation, the particles were allowed to move, but the fields were kept fixed (i.e., static and, of course, not self-consistent anymore). The evolution was traced over fourteen plasma times -- long enough for the particles to become substantially isotropized, so that the number of trapped particles diminished drastically. The radiation spectrum obtained in this run is entirely consistent with the synchrotron spectrum from the particles with a thermal spread, as is evident from Figure \ref{frozen}.

\section{Conclusions}

The primary results of this paper are as follows. First, we present a general expression for the spectral energy per solid angle emitted by an ensemble of particles in the small-angle jitter approximation, Eq. (\ref{dWmain}). Second, we have found that the electrons streaming through the filaments and being trapped in them produce a transient hard spectrum,  Eqs. (\ref{Omegab},\ref{trapjitt}). Third, we analyzed the large-angle deflection regime and showed that the spectrum starts to resemble the synchrotron spectrum near the peak, but a new spectral break at a lower frequency appears, Eq. (\ref{largejitt}). From the positions of the spectral peak and break, one can deduce the field correlation length. 

Fourth, PIC simulations show that the radiation spectrum produced at the onset of and during the phase of the exponential growth of the magnetic field is grossly inconsistent with synchrotron, Fig. \ref{spec}. The appearance of such a spectrum in the beginning of an emission episode can be used as a benchmark signal of the onset of the magnetic field generation in astrophysical sources and laboratory experiments. Among possible astrophysical systems where such emission can be or could have already been observed are gamma-ray bursts. The data show hard synchrotron-violating spectra in some bursts and the majority of spectra are flat \citep{Preece+98,Preece+00,Kaneko+06}, which are difficult to explain within the synchrotron model. Such spectra have recently been interpreted in the jitter emission paradigm \citep{MPR09}. We should caution the reader that ``blind'' application of the jitter spectrum template for interpretation of observational data, without checking the physical conditions at and the validity of the jitter approximation for the source in hand (galactic and quasar jets, supernovae remnants, etc.), can yield incorrect results. Possible laboratory experiments include laser-plasma interactions in which a beam (e.g., a probe electron beam) propagating through turbulent fields can emit jitter radiation \citep{RM11}.

Fifth, although the spectra after saturation are consistent with the synchrotron `template', an overall trend of the system toward the small-angle jitter regime (i.e., toward $\delta_{\rm jitt}<1$)  is observed in Fig. \ref{delta}, which suggest the scaling: $\delta_{\rm jitt}\propto\sqrt{\epsilon_B}\,\Gamma (\omega_p t)^{-0.5}$. Although the magnetic field decays, it does so rather slowly, $B(t)\propto (\omega_p t)^{-1}$, see Fig. \ref{epsB} (for example, for $\Gamma=10$ the systen shall return to the small-angle jitter regime at times $t\ga100$). So, if the field is continuously produced (as in the case of the propagating shock, for instance) the field decay can be compensated by the increase of the emitting volume, so that the total spectral emissivity will increase logarithmically, $P_{\rm tot}(\omega)\propto\int B\,dV\propto \ln(\omega_p t)\propto \ln(\omega_p L_{ps}/c)$, assuming these scalings hold at asymptotically late times, where $L_{ps}$ is the size (longitutinal extent) of the post-shock medium. Since the plasma time $\sim\omega_p^{-1}$ is generally very short in astrophysical sources, the overall time-integrated (and even time resolved, but with a coarse temporal resolution) spectrum can be dominated by the small-angle jitter spectrum, which is expected to be flat (unless the strong anisotropy of the fields and/or particles is somehow maintained). 

Here we also comment on the relation of our results and the simulations of \citep{SS09}. First of all, one should understand that the physical set-ups are entirely different. In this paper we study the evolution of radiation during the filamentation instability: the plasma in the simulation box is initially homogeneous, the particle distribution is unstable and the entire system is intrinsically non-stationary. In contrast, \citet{SS09} simulated a well-developed collisionless shock: a steady state system with very slow, if any, evolution of conditions and radiation spectra. Thus, the two simulations are complementary and the direct comparison of them is not well-posed. Since, however, the shock is moving through a medium with a constant speed, the temporal evolution of the filamentation instability and its subsequent saturation and further nonlinear evolution studied here is, to a certain degree, represented by the spatial profiles in the pre-shock and post-shock domains. The filamentation instability and its saturation occur far in front of the shock (hundreds or thousands skin lengths, in typical simulations) by particles escaping from the shock, hence this region would roughly correspond to the early and saturation times in our simulations. After saturation, mergers of magnetic filaments increase their sizes and radiation spectrum now mimics synchrotron, as our simulations show. In shock simulations, this merging stage occurs in a large region in front of the shock and this is where the radiation is collected in \citep{SS09}. Their results are, thus, in agreement with the ones presented here. The shock itself and the medium just behind the shock do not correspond to our simulations. However, the magnetic field strength decreases behind the shock and in a few hundred skin lengths radiation should enter the jitter regime. Unfortunately, radiation from neither the early pre-shock, nor from the far downstream regions have been shown in \citep{SS09}. We stress that radiation from the far downstream can be of great importance and dominate the entire shock emission, provided the scalings presented in the previous paragraph hold.

\acknowledgements
MVM work has been supported by NSF grant AST-0708213, NASA ATFP grant NNX-08AL39G and DOE grant  DE-FG02-07ER54940. MVM also acknowledges support from The Ambrose Monell Foundation (IAS) and The Ib Henriksen Foundation (NBIA). Computer time was provided by the Danish Center for Scientific Computing (DCSC). 


\end{document}